\documentclass[aps,prl,showpacs,showkeys,twocolumn,10pt,a4paper,superscriptaddress]{revtex4-1}

\usepackage{graphicx}
\usepackage{dcolumn} 
\usepackage{bm}
\usepackage{xcolor} 
\usepackage[caption=false]{subfig}
\usepackage{epstopdf}

\begin{document}

\title{Wetting heterogeneities in porous media control flow
  dissipation}

\author{Julie Murison} \email{julie.murison@ds.mpg.de}
\affiliation{Max Planck Institute for Dynamics and Self Organization,
  37077 G\"ottingen Germany}

\author{Beno\^{i}t Semin}%
\affiliation{Laboratoire de Physique Statistique, Ecole Normale
  Superi\'{e}ure, UPMC Univ Paris 06, Universit\'e Paris Diderot,
  CNRS, 24 rue Lhomond, 75005 Paris, France} \affiliation{Max Planck
  Institute for Dynamics and Self Organization, 37077 G\"ottingen
  Germany}

\author{Jean-Christophe Baret} \affiliation{Max Planck Institute for
  Dynamics and Self Organization, 37077 G\"ottingen Germany}

\author{Stephan Herminghaus} \affiliation{Max Planck Institute for
  Dynamics and Self Organization, 37077 G\"ottingen Germany}

\author{Matthias Schr\"oter} \email{matthias.schroeter@ds.mpg.de}
\affiliation{Max Planck Institute for Dynamics and Self Organization,
  37077 G\"ottingen Germany}

\author{Martin Brinkmann} \affiliation{Experimental Physics, Saarland
  University, 66123 Saarbr{\"u}cken, Germany} \affiliation{Max Planck
  Institute for Dynamics and Self Organization, 37077 G\"ottingen
  Germany}
	  
\date{\today}

%%%%%%%%%%%%%%%%%%%%%%%%%%%%%%%%%%%%%%%%%%%%%%%%%%
\begin{abstract}
  Pressure controlled displacement of an oil/water interface is
    studied in dense packings of functionalized glass beads with
    well-defined spatial wettability correlations. An enhanced
    dissipation is observed if the typical extension $\xi$ of the
    same-type wetting domains is smaller than the average bead
    diameter $d$. Three dimensional imaging using X-ray
    microtomography shows that the frequency $n(s)$ of residual
    droplet volumes $s$ for different $\xi$ collapse onto the same
    curve. This indicates that the additional dissipation for small
    $\xi$ is due to contact line pinning rather than an increase of
    capillary break-up/coalescence events.
\end{abstract}
%%%%%%%%%%%%%%%%%%%%%%%%%%%%%%%%%%%%%%%%%%%%%%%%%%


\maketitle

Flow of immiscible fluids in porous solids is encountered in nature
and in many technological applications ranging from groundwater or oil
reservoirs \cite{Dullien1991, Sahimi2011,Anderson1987a}, soils,
filtration membranes \cite{Kota2012, Shannon2008}, fuel cells
\cite{He2000,Sinha2007} or microfluidic systems \cite{Quake2005,
  Champagne2010, Cottin2011, Tabeling2012}. In all of these instances,
pore geometry and wall wettability are essential factors that control
the dynamics of slow interfacial displacement
\cite{Dullien1991,Sahimi2011}. Natural porous media such as rocks and
soils typically exhibit a large variability in wettability linked to
chemical heterogeneities of the rock and pore space
\cite{Nordgard2009,Hassenkam2009}. A detailed understanding of the
displacement processes down to the pore scale is important to optimize
CO$_2$ sequestration \cite{Bickle2009}, secondary and tertiary oil
recovery \cite{Sahimi2011}, and to design novel materials for
filtering, energy conversion, and fuel cells
\cite{He2000,Sinha2007}. Statistical descriptions which incorporate
pore scale geometry \cite{Balhoff2008,Battiato2011} into macroscopic
single phase flow descriptions have been able to largely improve
predictions of dissolution and chemical transport through porous
media. The impact of heterogeneously wettable pore walls, however,
studied so far only in natural rock or restored samples
\cite{Kumar2010}.

In the well studied case of bead packs consisting of mixtures of
monodisperse wetting and non-wetting spheres, it has been found that
sample scale wettability can be described by averages of surface
energies over several pores \cite{Ustohal1998,OCarroll2005}. However,
experimental and numerical studies of mixtures of wetting and
non-wetting beads of different diameters indicated that the typical
size of wetting domains should play a significant role in fluid
invasion processes \cite{Sorbie2000,Fatt1959, Hazlett1998}. These
seemingly contradictory findings about the role of wetting
heterogeneities on different length scales need to be resolved in
order to formulate a predictive continuum theory of multiphase flows
in porous media\cite{Miller1998,Sahimi2011}.

In this Letter we show experimentally that the typical extension and
spatial correlation of differently wettable surfaces in a porous
medium strongly influence the displacement of immiscible
fluids. Random packings of monodisperse glass beads with suitably
functionalized surfaces allow us to prepare model porous media with
same-type wettability domains of typical size $\xi$. We choose
monodisperse bead packs in order to isolate effects caused by
differences in wettability from those caused by local pore
geometry. Capillary pressure saturation (CPS) experiments are used to
characterize global sample wettability in core flood experiments and
many studies synthetic porous media \cite{Anderson1987a}. Our
experiments establish a link between CPS curves and pore scale wetting
conditions.

%%%%%%%%%%%%%%%%%%%%%%%%%%%%%%%%%%%%%%%%
\begin{figure}[h!]
  \includegraphics[width=\columnwidth]{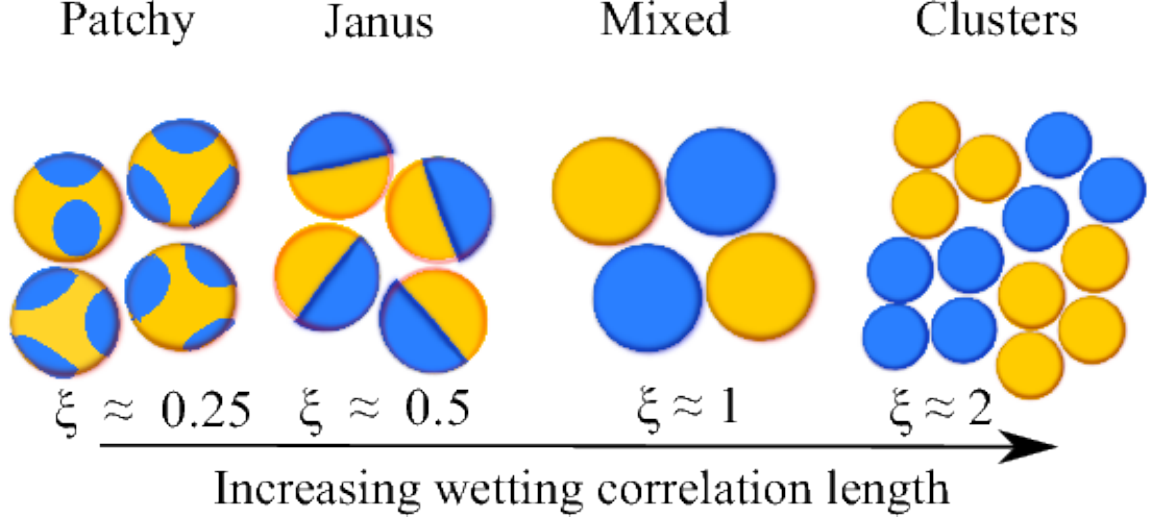}
  \caption{Sketch of the four model porous media considered. Oil
      and water wet surfaces are shown in orange and blue,
      respectively. All samples exhibit the same bead diameter $d$ and
      average surface energy, but differ in the typical extension
      $\xi$ (normalised by $d$) of the water wet domains.}
    \label{fig1}
\end{figure}
%%%%%%%%%%%%%%%%%%%%%%%%%%%%%%%%%%%%%%%%

%%%%%%%%%%%%%%%%%%%%%%%%%%%%%%%%%%%%%%%%
\begin{figure*}[floatfix]
  \includegraphics[width=\textwidth]{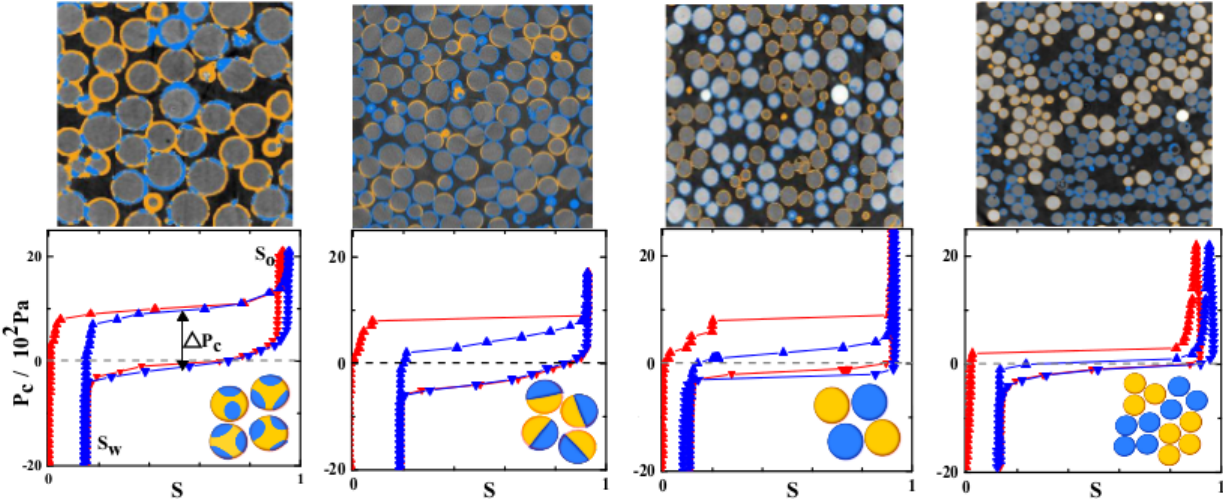}
  \caption{Representative CPS curves for Patchy, Janus, Mixed and
    Cluster samples. The hysteresis loop opening, $\Delta P_{\mathrm
      c}$, as well as the irreducible water ($S_{\mathrm w}$) and oil
    ($S_{\mathrm o}$) saturations are indicated on the first graph.
    X-ray microtomography of dry samples are shown above each
    curve. The blue highlighted surfaces represent the water wetting
    and the orange the oil wetting domains. From these images the
    correlation $\xi$ is calculated.}
  \label{fig2}
\end{figure*}
%%%%%%%%%%%%%%%%%%%%%%%%%%%%%%%%%%%%%%%%

In order to relate the CPS measurements to the pore scale
  processes we employ X-ray microtomography to detect the distribution
  of fluid phases and to monitor the overall shape of the invading
  front. So far, studies using three dimensional imaging of
immiscible two phase flows based on either X-ray microtomography
\cite{Berg2013, Ott2013} or confocal microscopy \cite{Datta2013} have
focused on the role of geometric heterogeneity \cite{Sahimi2011}. In
these samples percolation theory describes the droplet size
distributions of the residual phase \cite{Iglauer2010, Iglauer2012,
  Ott2013}. Our model systems allows us to test the validity of these
predictions in the presence of defined wetting heterogeneity.

Different protocols to functionalize the glass beads and to pack
  them allows us to create model porous media with wetting
  heterogeneities of different spatial extensions. As shown in
  Fig.~\ref{fig1}, the size of the same-type wettability domains $\xi$
  ranges from smaller than a single pore containing several different
  wetting patches up to multiple bead diameters $d$. In all samples
we control the coverage such that $(50\pm5)\%$ of the surface is oil
wet and $(50\mp5)\%$ water wet. The typical size $\xi$ of the surface
domains is quantified by the correlation length of the wettability
(see Supplemental Material). All samples consist of packed
  spherical glass beads (MoSci) with average diameter
  d=(375$\pm$~25)$\mu$m porosities in the range 0.38--0.40 where the
  untreated surface of the glass beads (cf.~Supplemental
  Material) constitutes the hydrophilic domains. Two different
coatings that modify the local surface energy are employed to create
the hydrophobic domains (gold, and CTMS). In the following, the beads
will be referred to according to their coating, i.e.~`gold' beads
refer to glass beads coated with gold. Gold beads are first sputtered
with gold (CreaVac, Dresden, thickness $\sim 200$nm) and then rendered
hydrophobic by a monolayer of hexadecanethiol (Sigma Aldrich). The
other coating `CTMS' consists of a Chlorotrimethoxysilane (CTMS, Sigma
Aldrich) layer directly coated on the glass bead. Packs of both gold
and CTMS coated beads exhibit hydrophobic behavior (see Supplemental
Material).
  
To design the samples displayed in Fig.~\ref{fig1} we used the
following protocols: (i) {\it Patchy} beads with $\xi \approx$ 0.25
created by masking the contact points in a granular pile and modifying
only the exposed surface; (ii) {\it Janus} beads with $\xi\approx$ 0.5
with one hydrophobic and one hydrophilic hemisphere; (iii) {\it Mixed}
samples with $\xi\approx$ 1 consisting of equal weight mixtures of
homogeneously wettable beads, and (iv) {\it Cluster} samples with $\xi
\approx$ 2 of equal weight mixtures of clusters $\approx 30$ of
spheres. Further details are provided in the supplementary materials.

Capillary Pressure Saturation (CPS) characteristics of each sample
were measured in a self-built experimental set-up described in the
supplementary information. The bead pack in the cell is bounded by two
membranes that are permeable for only one of the two fluid
phases. During an CPS experiment the pressure difference $P_{\mathrm
  c}=P_{\mathrm w}-P_{\mathrm o}$ between the fluid domains connected
to their respective reservoirs is slowly ramped up and down by changes
in the hydrostatic pressure difference. The mass of the displaced or
reinvading oil is measured as a function of time for a given value of
$P_{\mathrm c}$. Once a steady state is reached the value of the water
saturation $S$ is recorded and $P_{\mathrm c}$ is increased,
respectively, decreased in steps of 1 mbar. Figure~\ref{fig2} displays
the so acquired CPS curves. After primary oil displacement the CPS
curves of subsequent water and oil displacements fall onto the same
closed loop. The irreducible oil and water saturations,
  $S_{\mathrm o}$ and $S_{\mathrm w}$, respectively, here plotted in
  terms of the water saturation $S$ display no systematic dependence
  on $\xi$ (see supplementary information).

%%%%%%%%%%%%%%%%%%%%%%%%%%%%%%%%%%%%%%%%
\begin{figure}[!ht]
  \center
  \includegraphics[width=\columnwidth]{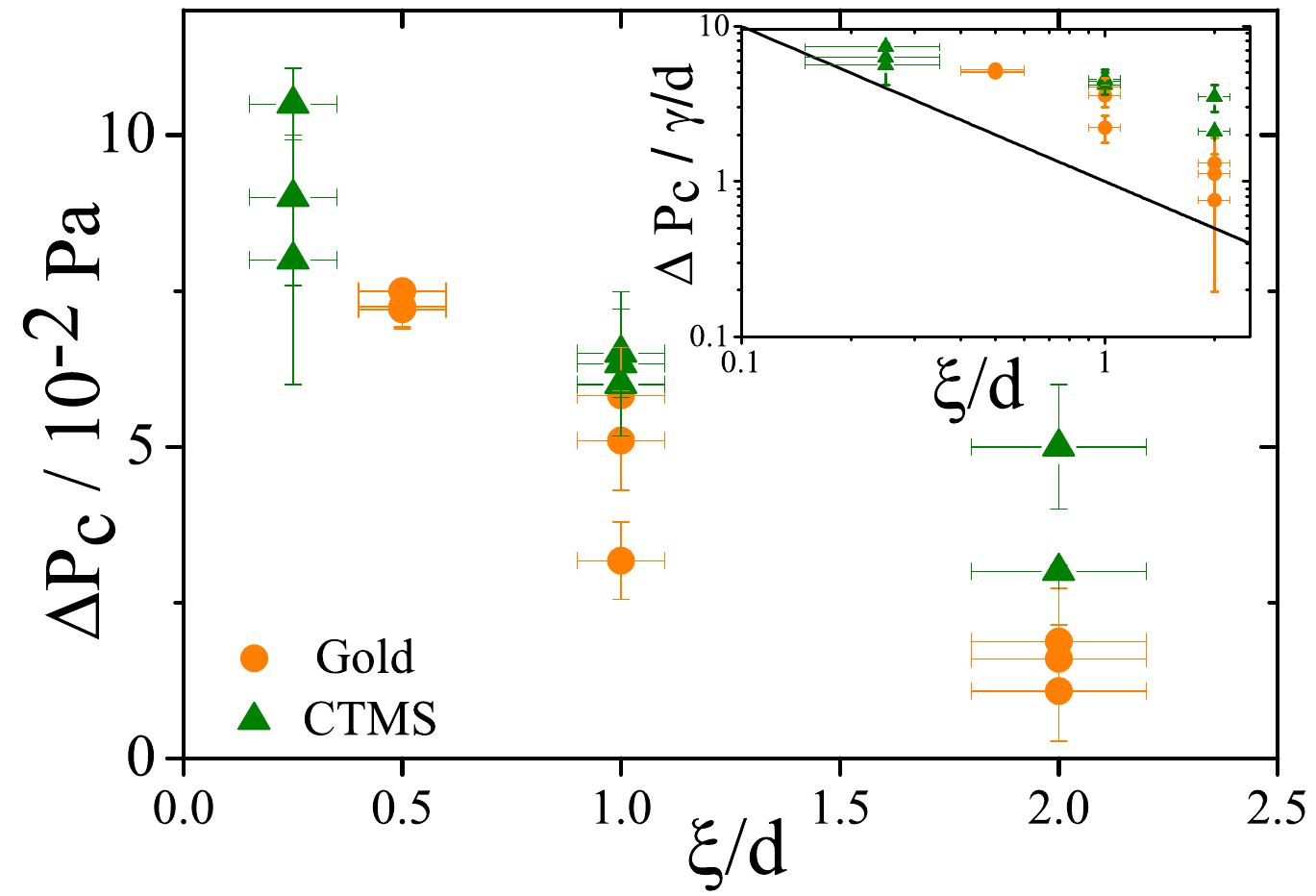}
  \caption{Hysteresis loop opening $\Delta P_c$ for different
      sample types and coatings against the ratio of wetting domain
      size $\xi$ to bead diameter $d$. Error bars of $\Delta P_c$
      indicate variations between successive cycles. Inset: same data
      in comparison to the ratio of interfacial tension $\gamma$ and
      bead diameter $d$.}
  \label{fig3}
\end{figure}
%%%%%%%%%%%%%%%%%%%%%%%%%%%%%%%%%%%%%%%%

In contrast, the typical size of the same type wetting domains,
  $\xi$, has a direct influence on the hysteresis loop opening $\Delta
  P_{\mathrm c}$ defined as the distance between the upper (water
  invasion) and lower (oil invasion) branch at $S=(S_{\mathrm
    o}+S_{\mathrm w})/2$. The data plotted in Fig.~\ref{fig3} show
  that the magnitude of $\Delta P_{\mathrm c}$ monotonously decreases
  with increasing $\xi$ for both types of coating (Gold and
  CTMS). Experiments with homogeneous water and oil wet bead packs
  yield values for $\Delta P_{\mathrm c}$ on the same order as for the
  Mixed and Cluster samples $(4 \pm 1)\cdot 10^2$Pa.  As the inset of
Fig.~\ref{fig3} shows, the obvious suggestion that $\Delta P_{\mathrm
  c}$ might be given by $\gamma/\xi$ (solid line) accounts for the
rough trend, but does not describe the data appropriately.

The area enclosed by the hysteresis loop $~\Delta P_{\mathrm
  c}(S_{\mathrm o}-S_{\mathrm w})$ is a direct measure of the work
dissipated during fluid displacement. Capillary hysteresis is a
permanent hysteresis, i.e.~the dissipation is still present in the
limit of a quasi-static interfacial advance. As $S_{\mathrm
    o}-S_{\mathrm w}$ is virtually independent on $\xi$ the area of the
  stable loop decreases when increasing $\xi$.  The additional
dissipation observed for small $\xi$ can be understood from the
pinning and depinning of three phase contact lines at the boundaries
of water and oil wet surface domains. These pinning sites are absent
in samples with $\xi>d$ while for samples with $\xi<d$, the density of
pinning sites increases as $\xi$ decreases.

Our hypothesis that contact line pinning causes the enhanced capillary
hysteresis for $\xi<d$ is further supported by the indication that the
topology of the wetting domains affects interfacial
displacement. The CPS curve for the Patchy sample in
Fig.~\ref{fig3} is centered around a higher pressure compared to
the other samples. Consistent with all the samples, the surfaces
of these beads are 50\% oil and 50\% water wet, however, the
hydrophobic part is a connected surface whereas the hydrophilic
portion exists as discrete patches. It appears that this topological
feature results in an asymmetry of apparent wettability, similar
to experimental observations made on planar surfaces with
analogous features \cite{Priest2007}.
    
In order to differentiate the additional dissipation in the presence
of wetting heterogeneities caused by pinning/depinning of the three
phase contact line from other mechanisms we employed in situ
X-ray microtomography imaging of the fluid distributions with a
$17\mu$m resolution. By comparing the size distribution of
disconnected droplets for samples with different correlation length,
we can test if capillary break-up/coalescence events contribute to the
differences in observed hysteresis. Technical details on the image
acquisition and analysis are presented in the supplementary
materials. 

%%%%%%%%%%%%%%%
%To supp matt
%%%%%%%%%%%%%%%
% A solution of 2M KI$_{(aq)}$ \mb{as the aqueous phase was used} to
% achieve X-ray contrast between the phases and images of the liquid
% distributions were made with a resolution of $17\mu$m per voxel.

%%%%%%%%%%%%%%%%%%%%%%%%%%%%%%%%%%%%%%%%%
\begin{figure}[ht!]
  \includegraphics[width=\columnwidth]{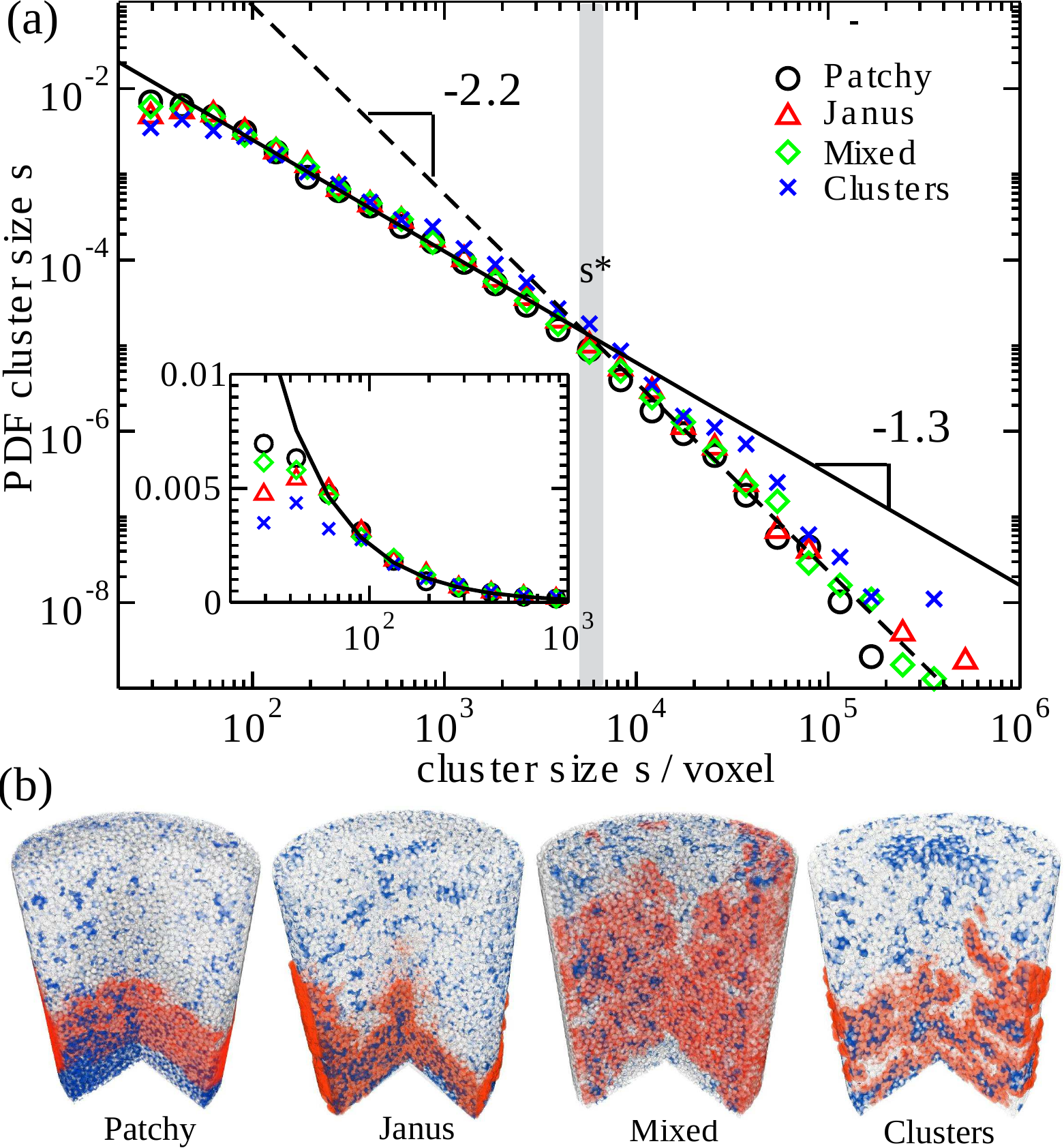}
  \caption{(a) Normed frequency $n(s)$ of droplets of volume $s$
      for trapped water in partially re-invaded bead packs from X-ray
      tomography after logarithmic binning \cite{Newman2005}. The
      voxel size is $17\mu$m. (b) Aqueous phase (blue) and active
      front (red) inside the different bead packs imaged during the
      second aqueous phase invasion, at a saturation of
      $S\approx0.3$.}
  \label{fig4}
\end{figure}
%%%%%%%%%%%%%%%%%%%%%%%%%%%%%%%%%%%%%%%%

A statistical analysis of the X-ray tomography images reveals only
small qualitative differences in the size distribution of the residual
phase between different samples and measurements at different points
of CPS curves. Images taken at saturations $S=S_{\mathrm o,w}$
  revealed very broad distributions of droplet sizes of the residual
  phases. As typical for fluid invasion processes, the majority of
the residual phase is contained in a single domain whose volume and
morphology may vary between different cycles of the experiments
explaining observed experimental scatter in $S_{\mathrm o}$ and
$S_{\mathrm w}$.\cite{Iglauer2010, Iglauer2012,
    Ott2013}

Apart for very small droplet volumes, the volume distribution of the
residual phase shown in Fig.~\ref{fig4}(a) follows a power law
$n(s)\propto s^{-\tau}$ with an exponent of $\tau = $1.3$\pm $0.1 for
volumes $s \lesssim s^\ast\approx 5000$ voxel while for $s>s^\ast$,
the distribution is consistent with the exponent $\tau \approx
2.2$. Both power-law scalings are robust towards image
processing. This was rigorously tested by applying morphological
filters (erosion, dilation, opening, closing) which locally change the
connectivity of small droplets in order to overestimate potential
effects caused by segmentation. The cross-over volume, $s^*$, compares
well with a lower estimate of the average pore volume $\approx 3000$
voxels from a Delaunay tessellation of a random bead packing (radius
$R \approx 10$ voxel, packing fraction $\phi \approx 0.6$, average
coordination number $N_c \approx 6$). As residual droplet of
  volumes $s\gtrsim s^\ast$ extend of several neighboring pores, their
  formation and frequency can be explained by invasion percolation
  with trapping predicting $\tau=2.18$ \cite{Sahimi2011}. Similar
  values of $\tau $ have been recently reported for immiscible fluid
  displacements in natural sandstones \cite{Iglauer2010, Iglauer2012,
    Ott2013}.

The insensitivity of the volume distributions $n(s)$ with respect to
$\xi$ suggests that wetting heterogeneities do not favor an
increase of capillary breakup and/or coalescence events of fluid
interface which are a source of dissipation in the displacement
processes. Hence, we can conclude that the main contribution to the
additional dissipation observed for $\xi<d$ is very likely caused
by irreversible pinning/depinning events of the three phase contact
line at boundaries between differently wettable surface domains.

To assess the influence of the wetting heterogeneities on the
  invasion process on the sample scale we also considered the overall
  morphology of the residual phase and invading fluid. Figure
  \ref{fig4}(b) shows the location of the `active' interface between
  aqueous and oil phase connected to their respective reservoirs and
  domains of the aqueous phase. We find the overall shape of the
  active interface to be more compact in Patchy and Janus samples, in
  contrast to the Mixed and Cluster samples which display a higher
  level of fingering (cf.~supplementary information). Qualitatively,
  the same observation is made on both branches of the hysteresis
  loop. Differences in the shape of the active interface at a given
  saturation is apparent only between the first and second water
  invasion while no significant difference between successive cycles
  can be detected.

In summary, we have shown that the pore scale and sub-pore scale
  wettability distribution in a porous media has a profound effect on
  capillary hysteresis and dissipation during immiscible fluid
  displacement. Besides droplets with dimensions well below the pore
  scale, the volume distribution of residual fluid droplets is barely
  affected by the length scale of the wetting heterogeneity and still
  consistent with percolation theory. Future work employing time
  resolved X-ray microtomography of interfacial advance and contact
  line pinning in small ensembles of pores could provide helpful
  insights into the dissipation mechanisms, emerging droplet distributions and front
  morphologies.

%%%%%%%%%%%%%%%%%%%%%%%%%%%%%%%%%%%%%%%%%%%%%%%%%%%%%%%%%%%%%%%%%%%%%%

{\it Acknowledgements,--} We gratefully acknowledge Thomas Hiller and
Daniel Herde for help with the correlation length calculations, Fabian
Schaller, Mario Scheel and Renaud Dufour for supporting measurements
and scientific discussions and Wolf Keiderling and Udo Krafft for
building the experimental cells. We also acknowledge Steffen Berg for
fruitful discussions regarding the droplet size distribution
analysis. We acknowledge generous support from BP Exploration
Operation Company Ltd. within the GeoMorph research project.

%%%%%%%%%%%%%%%%%%%%%%%%%%%%%%%%%%%%%%%%%%%%%%%%%%%%%%%%%%%%%%%%%%%%%%

%\bibliographystyle{apsrev}
%\bibliography{references_140317}

\pagebreak
\clearpage

\end{document}